\documentclass[superscriptaddress,showkeys]{revtex4} 
\usepackage{graphicx,amsmath,latexsym,amssymb}
\usepackage{color} 
  
\usepackage[hidelinks]{hyperref}
\hypersetup{
  colorlinks   = true, 	
  urlcolor     = blue, 	
  linkcolor    = blue, 	
  citecolor   = magenta 	
}

\usepackage{orcidlink}

\usepackage{lineno}
\usepackage{bm}
\newcommand{\mathsym}[1]{{}}

\usepackage{graphicx}
\newcommand{\be}{\begin{equation}}
\newcommand{\ee}{\end{equation}}
\newcommand{\bea}{\begin{eqnarray}}
\newcommand{\eea}{\end{eqnarray}}
\newcommand{\bd}{\begin{displaymath}}
\newcommand{\ed}{\end{displaymath}}
\newcommand{\ad }{a^{\dagger}}
\newcommand{\p }{ \partial}
\newcommand{\de }{\nu }
\newcommand{\lb }{ \left (}
\newcommand{\rb }{ \right )}

\begin{document}

\title{Parity-deformed  $sl(2,R)$, $su(2)$ and $so(3)$ Algebras: a Basis for Quantum Optics and Quantum Communications Applications}

\author{W. S. Chung\orcidlink{0000-0002-1358-6384}}
\email{mimip44@naver.com}
\affiliation{Department of Physics and Research Institute of Natural Science,
 College of Natural Science, Gyeongsang National University, Jinju 660-701, Korea}

\author{H. Hassanabadi\orcidlink{0000-0001-7487-6898}}
\email{hha1349@gmail.com}
\affiliation{Departamento de F\'{\i}sica Te\'{o}rica, At\'{o}mica y \'{O}ptica, and Laboratory for Disruptive Interdisciplinary Science (LaDIS), Universidad de Valladolid, 47011 Valladolid, Spain}

\author{L. M. Nieto\orcidlink{0000-0002-2849-2647}}
\email{luismiguel.nieto.calzada@uva.es}
\thanks{Corresponding author}
\affiliation{Departamento de F\'{\i}sica Te\'{o}rica, At\'{o}mica y \'{O}ptica, and Laboratory for Disruptive Interdisciplinary Science (LaDIS), Universidad de Valladolid, 47011 Valladolid, Spain}

\author{S. Zarrinkamar\orcidlink{0000-0001-9128-4624}}
\email{saber.zarrinkamar@uva.es}
\affiliation{Departamento de F\'{\i}sica Te\'{o}rica, At\'{o}mica y \'{O}ptica, and Laboratory for Disruptive Interdisciplinary Science (LaDIS), Universidad de Valladolid, 47011 Valladolid, Spain}
\affiliation{Department of Basic Sciences, Garmsar Branch, Islamic Azad University, Garmsar, Iran}

\date{\today}

\begin{abstract}
Having in mind the significance of parity (reflection) in various areas of physics, the single-mode and two-mode Wigner algebras are considered adding to them a reflection operator. 
The associated  deformed $sl(2, R)$ algebra, $sl_{\nu}(2,R)$ and the deformed $so(3)$ algebra, $so_{\nu}(3)$, are constructed for the widely used Jordan-Schwinger and Holstein-Primakoff realizations, commenting on various aspects and ingredients of the formalism for both single-mode and two-mode cases. Finally, due to its potential application in the study of qubit and qutrit systems, the parity-deformed $so_{\nu}(3)$ representation is analyzed based on the isomorphy of $so(3)$ and $su(2)$. Related applications are discussed as well.
\end{abstract}


\maketitle

\section{Introduction}

After a century of the advent of quantum mechanics, this theory still has outstanding challenges and open questions to debate in many basic concepts and formulas of the theory. As a result, a variety of modifications to the related fundamental relationships have been proposed, primarily in an attempt to investigate not only  problems in physics, but also in other domains of science and technology beyond our current experimental techniques and facilities. 
On the other hand, it is well known that theoretical physics has always been inspiring and that the history of physics is full of 
examples of theoretical predictions prior to their experimental verification. This makes theoretical and mathematical physics continue to be an attractive field of study for the scientific community, competing with the incredible technological achievements that are being achieved in recent times. 

Given the current great technological importance of quantum theory in modern technologies, in what is called {\it second quantum revolution}, which covers applications in various fields, such as quantum information, quantum communication, quantum teleportation and quantum technologies in general, we are interested in analyzing  possible generalizations of some fundamental basic results that can be contrasted using current or future experimental techniques in development.
 In particular, remembering here the importance that the analysis of the harmonic oscillator problem has had in both optics and quantum technologies, we are going to theoretically consider a generalization of Wigner's algebra modified by parity inclusion. 
 To be precise, what is considered here is a kind of parity-warped operator algebra, sometimes called Wigner, Dunkl, or Wigner-Dunkl algebra.
 The pillars of the formulation of the problem and the consideration of the reflection operator go back to the works of Holstein and Primakoff, Wigner, Baxter and later  Dunkl \cite{Wigner,Yang,Raiford,Dunkl}. Generalization of algebra is usually done by the Jordan-Schwinger and Holstein-Primakoff realizations of the associated creation and annihilation operators and their extensions \cite{Holstein, Arik,Biedenharn,Macfarlane, Palev, Horvathy}. In physics, however, the applications are quite interesting and include a variety of concepts such as coherent states, compressed states, cat states, fidelity, entanglement transfer, etc. Let us now review, in chronological order, some of the most notable works that are closely related to the objectives of our work.

In pioneering work, with a goal quite different from that pursued here, Plyushchay investigated the deformed Heisenberg algebra with reflection in \cite{Plyushchay}, demonstrating that the normalized deformed algebra has the form of a  guon algebra \cite {guon, guon 2}.
Using a reflection-deformed Heisenberg algebra, a unified description of anyons, bosons and fermions was analyzed in the instructive survey \cite{Horvathy 2010}. 
The hidden supersymmetry and related tri-supersymmetric structure of some physical examples were investigated in full detail  by Jakubsk\'y et al. \cite{Jakubsky}, who also studied the role of parity. In interesting novel work, Dehghani and others studied the Wigner-Heisenberg algebra in relation to so-called cat states \cite {Dehghani 2015}.
The same group later analyzed a parity-deformed version of the Jaynes-Cummings model, studying in detail the entangled states in a lossy cavity interacting with an external field \cite{Dehghani 2016}.  Also, a very interesting connection to generalized Rabi model has been proved by Moroz \cite {Moroz}.

The hidden superconformal symmetry of the harmonic oscillator problem was obtained by a non-local Foldy-Wouthuysen transformation  in \cite{Inzunza}. The consequences of parity deformation on entanglement transfer to distant atomic qubits were considered in \cite{Dehghani 2019}, where the qubits were assumed to be embedded in two lossy cavities connected by a leaky fiber.

In \cite{Sargolzaeipour 2019} a form of generalized Dunkl oscillator has been formulated in both relativistic and non-relativistic cases. 
Quantum error tracking with parity measurements was discussed in \cite{Mohseninia 2020}. The analysis of dynamical transformations between quantum reference frames, including the parity properties, was formulated by Ballesteros et al. \cite{Ballesteros 2021}. 
The optical properties of a pumped-dissipative quantum dot cavity with a parity-deformed Heisenberg algebra were investigated in \cite{Hurtado-Molina} and it was shown that the deformation produces a collective effect on the photonic emission of the system.  Fakhri et al. studied a two-photon parity-deformed Jaynes-Cummings model and discussed in detail the effect of deformation on important physical quantities, including fidelity \cite{Fakhri 2021}.
Fakhri and Sayyah-Fard \cite{Fakhri 2022} derived and discussed in depth the $x$ representations of the coherent and cat states of the para-bosonic oscillator algebra. The supersymmetric structure of deformed Heisenberg algebras with reflection has been studied in  \cite{plu frac susy 1996, plu 1996 hidden nonlinear ijmpa, correa ann 2007, plu plb 2007, Correa prl 2008, Dong 2022}. Also, the generalization of the Dunkl oscillator to Calogero–Sutherland model has been well addressed in \cite {artemio 1, artemio 2}.

 The effect of parity on quantum optimization and computation was well covered in \cite{Drieb 2023}, \cite{Fellner 2023} and \cite{Raussendorf 2023}.
The problem of two qubits moving in a parity-deformed field was considered in \cite {Mojaveri 2023}, where it was shown that parity warping can give us greater protection against entanglement. 
The gauge invariance of a charged Wigner-Dunkl system was analyzed in \cite{Junker 2023} and it was shown that the gauge invariance holds in case of minimal coupling to a vector potential. 
In reference \cite{Sedaghatnia 2023} a generalization of the Wigner-Dunkl problem was considered and the related coherent states were studied, as well as the corresponding Mandel parameter and the bunching/anti-bunching effect.  
Very instructive and quite up-to-date applications, as well as the importance of parity in the study of fidelity and quantum error correction, are investigated in references \cite {Zhu 2023, Lu 2023}.

In \cite{Buzek} the Jaynes-Cummings model with intensity-dependent coupling with a Holstein-Primakoff $SU(1,1)$ state was considered, and it was shown that the revivals of the radiation squeezing exhibit a periodic character. 
The generalization of the Jaynes-Cummings model was studied using the deformed harmonic oscillator problem and the Holstein-Primakoff realization of $su(1,1)$ by Chaichian and coworkers \cite{Chaichian}. 
The Holstein–Primakoff  transformation is also present in many other fields, including the laser-cooled trapped ion interaction \cite {Koroli}, the $N$-body Lipkin model \cite{Hirsch}, and equilibrium electron-magnon systems \cite{Bajpai 2021}. 
The transfer of quantum information between a nanomechanical resonator and a superconducting transmission line resonator via a charged qubit in the case of two non-degenerate modes was considered using the Jordan-Schwinger realization in \cite{Sun}. 
The generalization of the Jaynes-Cummings and Dicke models was formulated using a generalized Jordan–Shchwinger realization and therefore the associated integrability and superintegrability were discussed in \cite {Skrypnyk}. The realization of three-dimensional Jordan-Schwinger problems was studied in \cite{Sunil Kumar}.  
Finally let us mention that the $SO(3)$ group has been used to investigate modern applications of quantum theory, including the maximally entangled states of qubits \cite{Zola, Rau 2021} and qutrits \cite {Uskov}. The perfect transfer of path-entangled and single-photon states in the so-called ${J_x}$ photonic lattices was discussed in the novel paper by Perej-Leija and co-workers \cite {Perej-Leija 2013}.

In the present work, due to the wide application of parity-dependent systems and the already mentioned applications in quantum communication and computing, we aim to investigate the effect of parity in a special form of generalized Wigner algebra.

This article is organized as follows. In Section~\ref{2}, the single-mode Wigner algebra and its realizations are reviewed, while the two-mode Wigner algebra is discussed in Section~\ref{3}. 
In both cases, what will be called $\nu$-deformed Wigner algebra will be defined and studied.
Two realizations of the algebra $su_{\nu}(2)$, the Jordan-Schwinger realization and the Holstein-Primakoff realization, are investigated in Section~\ref{4}, while a representation of the $so_{\nu}(3)$ algebra is studied in Section~\ref{6}. Finally, in Section~\ref{7} we summarize the main conclusions of the present work.

\section{Single-mode Wigner Algebra and its realizations}\label{2}

The single mode Wigner algebra \cite {Holstein} is given by the commutation relations
\be\label{Walgebra}
[a, \ad] = 1 + 2 \nu R, \qquad  [N, \ad]=\ad, \qquad [N, a]=-a,
\ee
where $a$ and $\ad $ are annihilation and creation operators, respectively,  and $ \nu>-\frac{1}{2}$ with $\nu$ being the Wigner parameter. The reflection operator $R$ is defined by
\be
R= (-1)^N,
\ee
and obeys the following anticommutation relations
\be
\{ R, a \} = \{ \ad , R \} =0 , 
\ee
as well as
\be
 R^{\dagger} = R^{-1} = R, \qquad R^2 = I.
\ee
Before proceeding further, it should be noted that although the Fock-Bargmann is considered in the complex domain, Sifi and Soltani introduced a Hilbert space on which the Dunkl operator on the real line \cite {proof 2002, proof 2004}. Therefore we preserve the common notation $x$ here. Let us introduce the eigenvectors of the number operator $N$ as 
$
N |n\rangle = n |n\rangle$, $n = 0, 1, 2, \dots$ \cite{Sargolzaeipour 2019},
where $N$ is a Hermitian operator and the ground state $|0\rangle$ satisfies
$
a |0\rangle  = 0.
$
From \eqref{Walgebra}, we have the following representation  \cite{Plyushchay} 
\begin{align}
a |n\rangle = \sqrt{ [n]_{\nu}}\ |n-1\rangle,
\qquad
\ad |n\rangle = \sqrt{ [n+1]_{\nu}}\ |n+1\rangle,
\end{align}
where the $\nu $-deformed numbers are defined as
\be\label{deformednumbers}
[n]_{\nu} = n + \nu ( 1 - (-1)^n ).
\ee
In fact, the first $\nu $-deformed numbers are \cite{Sargolzaeipour 2019}
\be
[0]_{\nu}  = 0,  \qquad
[1]_{\nu}  = 1+2\nu,  \qquad
[2]_{\nu}  = 2,  \qquad
[3]_{\nu}  = 3+2\nu, \qquad
[4]_{\nu}  = 4.
\ee
In general, we have
\be
[2k]_{\nu} = 2k  , \qquad[2k+1]_{\nu} = 2k+1+2\nu , \qquad k=0,1, 2, \dots 
\ee
Some properties for the $\nu $-deformed numbers are the following
\begin{align}
[n]_{\nu}  + [n+1]_{\nu} &= 2n +1 + 2 \nu,
\nonumber  \\
[n+2]_{\nu}  - [n]_{\nu}  &=2,
\nonumber  \\
[n]_{\nu}  + [n+2k+1]_{\nu} &= 2n +2k+1 + 2 \nu, \quad
 k \in \mathbb{Z},
\nonumber  \\
[n+2k]_{\nu}  - [n]_{\nu}  &=2k, \quad k \in \mathbb{Z},
\nonumber  \\
[m]_{\nu} [n+1]_{\nu} - [n]_{\nu} [m+1]_{\nu}
&= m-n - \nu ( 2n+1)(-1)^m  + \nu (2m+1) (-1)^n - 2\nu^2 ((-1)^m - (-1)^n ) .
\label{15}
\end{align}
If we split the set of integers as $\mathbb{Z}= \mathbb{Z}_0\oplus \mathbb{Z}_1$, with $\mathbb{Z}_0=\{0, 2, 4, \dots\}$ and $\mathbb{Z}_1 = \{1, 3, 5, \dots \}$,  Equation \eqref{15} can be rewritten as
\be
[m]_{\nu} [n+1]_{\nu} - [n]_{\nu} [m+1]_{\nu} = \left\{
\begin{array}{ll}
(m-n)( 1 + 2 \nu), &  n \in \mathbb{Z}_0,  m \in \mathbb{Z}_0, \\
m-n + 2 \nu(m+n+1)+4\nu^2, &   n \in \mathbb{Z}_0,  m \in \mathbb{Z}_1, \\
m-n - 2 \nu(m+n+1)-4\nu^2, &  n \in \mathbb{Z}_1,  m \in \mathbb{Z}_0, \\
(m-n)( 1 - 2 \nu), &   n \in \mathbb{Z}_1,  m \in \mathbb{Z}_1 .
\end{array}
\right.
\ee
Now we will look for the coordinate realization for the single-mode Wigner algebra \eqref{Walgebra}. The realization depends on the choice of the basis functions. For the ordinary boson algebra, we have the realization \cite{book ladder}
\be
a = \p_x, ~~~~ \ad = x, ~~~ N =\ad a = x \p_x.
\ee
For the monomial basis given by $f_n (x)= x^n $ ($ n=0, 1, 2, \dots$), this realization is 
\be
af_n  (x)= n f_{n-1} (x), ~~~ \ad f_n  (x)= f_{n+1} (x), ~~~ N f_n  (x)= n f_n (x).
\ee
where the domain doesnot include the origin. Since $ a f_0=0$, this representation is bounded from below and $ f_0$ is called the ground state.
Considering now the so-called quasi-polynomial basis $\phi_n (x)$ given by
 \be
 \phi_n  (x)= \frac{ \Gamma (x+1)}{\Gamma ( x + 1 -n )}, ~~~ n=0, 1, 2, \dots,
 \ee
a new realization for the ordinary boson algebra \eqref{Walgebra} appears as
\be
 a = e^{\p_x} -1, ~~~ \ad = x e^{-\p_x}, ~~~ N= x \left( 1 - e^{-\p_x}\right)= \ad a,
 \ee
where  $\partial_x:=\frac{d}{dx}$ is the ordinary derivative, $e^{\pm\partial_x} f(x):=f(x\pm 1 )$ and the corresponding action is 
\be
a\phi_n  (x)= n \phi_{n-1} (x), ~~~ \ad \phi_n  (x)= \phi_{n+1} (x), ~~~ N \phi_n  (x)= n \phi_n (x).
\ee

\subsection{Monomial basis Wigner algebra realization}

For the monomial basis $ f_n (x)=x^n$, the realization of the Wigner algebra  \eqref{Walgebra} is
\be \label {finite}
a = \p_x + \frac1x \left( \de - \de (-1)^{x \p_x}\right), 
~~~~ \ad = x, ~~~ N = x \p_x, ~~~
R = (-1)^{x \p_x}.
\ee
The representation of this algebra in this basis is
\be
af_n  (x)= ( n +\de -\de (-1)^n ) f_{n-1} (x), ~~~ \ad f_n  (x)= f_{n+1} (x), ~~~ N f_n  (x)= n f_n (x).
\ee
It should be noted that  \eqref{finite} correspond to the algebra of finite-difference equations \cite {umbral, book finite difference, turbiner calogero, finite mpla 1995, turbiner jpa, Turbiner phys rep}

\subsection{ Quasi-polynomial basis Wigner algebra realization}

In this subsection we derive the quasi-polynomial basis for the Wigner algebra, which is related to $Z_2$-grading, which implies that $ R f_n  (x)= (-1)^n f_n  (x)$. Thus, we change the quasi polynomial for an ordinary boson algebra into the $Z_2$-graded quasi polynomial for the Wigner algebra.
The $Z_2$-graded quasi polynomial basis is defined as
\be
\label{Gaussian}
\phi^{\de}_n(x):=\prod_{k=0}^{n-1}(x-k - \de (-1)^k )
,\quad n\geq 1,
\ee
with $\phi^{\de}_0(x):=1$. Therefore, the realization is
\be
\label{loweringandraising}
a=T_{\de} e^{ \partial_x}-1, \qquad\ad = ( x -\de )T_{\de}e^{- \partial_x},
\ee
 where 
\bd
 T_{\de} F(\de) := (-1)^{\de \frac{d}{d \de}} F(\de) = F(-\de).
 \ed
One can easily check that the above realization obeys
\begin{align}
a\phi^{\de}_n(x) &=[n]_{\nu}\ \phi^{\de}_{n-1}(x),
\nonumber  \\
a^\dag \phi^{\de}_n(x)&=\phi^{\de}_{n+1}(x),
\nonumber  \\
\label{samaholl}
\phi^{\de}_n(x) &=(a^\dag)^n 1,
\\
[a,a^\dag] \phi^{\de}_{n}(x)
& = ( 1 + 2 \de R )\, \phi^{\de}_{n}(x),
\nonumber 
\end{align}
where
\be
N = \ad a-\de + \de R. 
\ee
Therefore,   the set of polynomials $\{{\phi}^{\de}_n(x)|n=0,1,\dots\}$
provides a basis for the realization of the Wigner algebra.

\section{Two-mode Wigner algebra}\label{3}

We can extend the single-mode Wigner oscillator algebra into the two-mode Wigner oscillator algebra as follows \cite{Raiford}
\begin{align}
[a_i, \ad_j] &= \delta_{ij} ( 1 + 2 \nu R_i), \qquad [a_i, a_j] =0, \qquad[\ad_i, \ad_j]=0,
\label{30-1}
\\
[N_i, \ad_j] &=\delta_{ij} \ad_i, \qquad\qquad\quad\ \ [N_i, a_j]=-\delta_{ij} a_i,
\label{30}
\end{align}
where $i=1,2$, and
\be
R_i = (-1)^{N_i},
\qquad
\{ R_i, a_i\} =\{\ad_i, R_i\} =0,
\qquad
[R_i, a_j] =[R_i, \ad_j]=0, ~~~\text{for} ~ i \ne j.
\ee
The number operators and the associated ladder operators satisfy \eqref{deformednumbers}
\be
\ad_i a_i = [N_i]_{\nu}. 
\ee
The eigenvectors of the number operators $N_i$ are defined as
\be
N_i |n_1, n_2\rangle = n_i |n_1, n_2\rangle ,  ~~~ n_i = 0, 1, 2, \dots,
\ee
where $N_i$ are  Hermitian, and the ground state $|0,0\rangle$ obeys
$
a_i |0, 0\rangle  = 0$,  $i=1,2$.
From \eqref{30}, we have
\begin{align}
a_1 |n_1, n_2 \rangle &= \sqrt{ [n_1]_{\nu}}\ |n_1-1, n_2 \rangle,
\quad
&\ad_1 |n_1, n_2\rangle = \sqrt{ [n_1+1]_{\nu}}\ |n_1+1, n_2\rangle,
\\
a_2 |n_1, n_2 \rangle &= \sqrt{ [n_2]_{\nu}}\ |n_1, n_2-1 \rangle,
\quad
&\ad_2 |n_1, n_2\rangle = \sqrt{ [n_2+1]_{\nu}}\ |n_1, n_2+1\rangle.
\end{align}
It is worth noting that the two-mode deformed Heisenberg algebra has been already discussed in connection with fractional helicity, Lorentz symmetry breaking, and anyons \cite {two-mode plu}.

\section{Realizations of  $sl(2,R)$ and $sl_{\nu}(2,R)$ Wigner algebras}\label{4}

We are going to consider two different realizations of the Wigner algebras  $sl(2,R)$ and $sl_{\nu}(2,R)$: the Jordan-Schwinger realization and the Holstein-Primakoff realization.

\subsection{Jordan-Schwinger realization based on the two-mode Wigner algebra}

From \eqref{30}, we have that the Jordan-Schwinger realization of the ordinary $sl(2,R)$ algebra is given by \cite {Sunil Kumar}
\be
\label{43}
J_+ = \ad_1 a_2, \qquad  J_- = a_1 \ad_2 , \qquad  J_0= \frac{1}{2} (N_1- N_2),
\ee
with
\be
[J_+, J_-]=2 J_0, \qquad  [J_0, J_{\pm}]= \pm J_{\pm}.
\ee
If we now consider the  two-mode Wigner oscillator algebra \eqref{30-1}--\eqref{30}, we will obviously need more generators to construct a Jordan-Schwinger realization due to the existence of reflection operators, and we can therefore write
\be
[J_+, J_-]=2 J_0 - \nu ( 2 N_2 +1 ) R_1 + \nu ( 2 N_1 +1 ) R_2 + 2 \nu^2 ( R_2 - R_1).
\ee
Therefore, if we now introduce the following three generators
\be
P = N_1 R_2 - N_2 R_1,
\qquad
K = \frac{1}{2}( R_2 - R_1),
\qquad
Q = \frac{1}{2}( R_2 + R_1),
\ee
then, the algebra $sl(2,R)$ plus the reflection symmetry (i.e., the algebra $sl_{\nu}(2,R)$) can be described with the above operators as follows
\begin{align}
[J_0, J_{\pm}] &= \pm J_{\pm},
\qquad\qquad \qquad\qquad\quad
[J_+, J_-] = 2 J_0  + 2 \nu P + 2 \nu ( 2 \nu +1 ) K,
\nonumber  \\
[K, Q] &= [K, P] = [K, J_0]=0, \qquad\,   \{ K , J_{\pm}\}=0,
\nonumber  \\
[Q, P] &= [ Q, J_0] =0, \qquad\qquad\qquad\ \{ Q, J_{\pm} \} =0,
\\
[P, J_0 ] &= 0, \qquad\qquad\qquad\qquad\qquad\ \ \{ P, J_{\pm} \} = \pm 2 Q J_{\pm} ~.
\nonumber  
\end{align}
The representation of the algebra is then given as
\begin{align}
J_0 |n_1, n_2 \rangle & = \frac { n_1 - n_2 }{2}\ |n_1, n_2 \rangle, 
\nonumber  \\
P|n_1, n_2 \rangle  &  =  ( n_1(-1)^{n_2 }- n_2 (-1)^{n_1 } ) |n_1, n_2 \rangle,
\nonumber  \\
Q|n_1, n_2 \rangle  &  =\frac{1}{2} \lb  (-1)^{ n_1} + (-1)^{ n_2} \rb\ |n_1, n_2 \rangle,
\nonumber  \\
K|n_1, n_2 \rangle   &  = \frac{1}{2} \lb  (-1)^{ n_2} - (-1)^{ n_1} \rb\ |n_1, n_2 \rangle,
\\
J_+ |n_1, n_2 \rangle  & = \sqrt{[n_1+1]_{\nu} [n_2]_{\nu}}\ |n_1+1, n_2-1 \rangle,
\nonumber  \\
J_-|n_1, n_2 \rangle  & =  \sqrt{[n_1]_{\nu} [n_2+1]_{\nu}}\ |n_1-1, n_2+1 \rangle,
\nonumber 
\end{align}
with $n_i = 0, 1, 2, \dots$
Introducing
\be
n_1 = j+m , \quad\qquad n_2 = j-m,
\ee
and
\be
 |\widetilde{j,m}\rangle= |j+m, j-m \rangle =|n_1, n_2 \rangle,
\ee
we have the following representation

\begin{align}
J_0  |\widetilde{j,m}\rangle   & = m\,  |\widetilde{j,m}\rangle,
\nonumber  \\
P  |\widetilde{j,m}\rangle   & = [ ( j +m )(-1)^{j-m} - (j-m)(-1)^{j+m} ]\  |\widetilde{j,m}\rangle,
\nonumber  \\
Q  |\widetilde{j,m}\rangle   & = \frac{1}{2} [(-1)^{j-m} + (-1)^{j+m}] \  |\widetilde{j,m}\rangle,
\nonumber  \\
K  |\widetilde{j,m}\rangle   & = \frac{1}{2} [(-1)^{j-m} - (-1)^{j+m}] \  |\widetilde{j,m}\rangle,
\\
J_+  |\widetilde{j,m}\rangle   & = \sqrt{[j+m+1]_{\nu} [j-m]_{\nu}} \  |\widetilde{j,m+1}\rangle,
\nonumber  \\
J_-  |\widetilde{j,m}\rangle   & = \sqrt{[j+m]_{\nu} [j-m+1]_{\nu}} \  |\widetilde{j,m-1}\rangle.
\nonumber 
\end{align}
This representation has dimension $(2j+1)$, which implies that the allowed value of $m$ are
\be
m = - j , -j+1 , \dots , j-1, j,
\qquad \text{with}\qquad
J_+ | \widetilde{j,j} \rangle = J_- | \widetilde{j,-j} \rangle = 0.
\ee
Therefore, the value of $2j$ is an integer.

\subsubsection{ Representation for odd $2j$}
In this case $2m$ is also odd. Thus, we have
\be
Q = 0, ~~~ K = R_J, ~~~ P = 2 j R_J,
\ee
where we set
$
R_J = (-1)^{ j - J_0}.
$
Thus, the $sl_{\nu}(2,R)$ algebra has three generators:
\begin{align}
[J_0, J_{\pm}]   & = \pm J_{\pm},
\qquad\quad 
[J_+, J_-]  =2 J_0  + 2 \nu  ( 2 \nu + j +1 ) R_J,
\\
R_J^2   & = I, \qquad\qquad\ \  [R_J, J_0]=0,  \qquad\quad\   \{ R_J, J_{\pm}\} =0.
\end{align}

\subsubsection{Representation for even $2j$}
In this case $2m$ is also even and we may write
\be
K = 0, \qquad\qquad Q = R_J, \qquad\qquad  P = 2 J_0 R_J.
\ee
Thus, the $sl_{\nu}(2,R)$ algebra has three generators satisfying
\be
[J_0, J_{\pm}]    = \pm J_{\pm},
\ \
[J_+, J_-]    =2 J_0 \lb 1 + 2 \nu
 R_J\rb,
\ \
R_J^2    = I, \ \   [R_J, J_0]=0,  \ \  \{ R_J, J_{\pm}\} =0.
\ee

\subsubsection{Examples}
\begin{itemize}
\item
{ $j=1/2$}:
In this case, we have the following matrix representation
\be
J_0 = \begin{pmatrix} 1/2 & 0  \cr 0 & -1/2
\end{pmatrix}, 
\qquad
J_+ = \begin{pmatrix} 0 & [1]_{\nu}   \cr 0 & 0
\end{pmatrix}, 
\qquad
J_- = \begin{pmatrix} 0 & 0  \cr [1]_{\nu}  & 0
\end{pmatrix}.
\ee
If we assign $ J_{\pm}$ to the $\nu$-deformed Pauli matrices $\sigma_{\pm}^{\nu}$, and preserve the third element in the same form of ordinary $\sigma_z$, i.e. considering $J_0$ as 
$\frac{1}{2} \sigma_z$, then we can  write
\be
[ \sigma_{+}^{\nu}, \sigma_{-}^{\nu}] = [1]_{\nu}^2 \sigma_{z}= ( 1 + 4 \nu + 4\nu^2 ) \sigma_{z},
\ee
where we used
$
R_J = 2 J_0.
$

\item
{ $j=1$}:
The matrix representation in this case is
\be
\hskip-0.5cm 
J_0 = \begin{pmatrix} 1 & 0 & 0 \cr 0 & 0 & 0 \cr 0 & 0 & -1
\end{pmatrix},
\
J_+ = \begin{pmatrix} 0 & \sqrt{[2]_{\nu}!} & 0 \cr 0 & 0 & \sqrt{[2]_{\nu}!} \cr 0 & 0 & 0 \end{pmatrix},
\
J_- = \begin{pmatrix} 0 & 0 & 0 \cr \sqrt{[2]_{\nu}!} & 0 & 0 \cr 0 & \sqrt{[2]_{\nu}!} & 0 \end{pmatrix}.
\ee
If we assign $J_0, J_{\pm}$ to the the $\nu$-deformed angular momentum generators  $L_z^{\nu}, L_{\pm}^{\nu}$, respectively, we have the following relationships
\be
\label{64}
[ L_{+}^{\nu}, L_{-}^{\nu}] =  2 L_{z}^{\nu} ( 1 + 2 \nu   L_{z}^{\nu} ),
\ee
where we used
$
R_J = L_{z}^{\nu}.
$
The algebra \eqref{64} is called quadratic algebra.

\item
{ $j=3/2$}:
Here, the representation takes the form
\begin{eqnarray}
\hskip-0.3cm 
J_+   = 
\begin{pmatrix} 0 & \sqrt{[3]_{\nu} [1]_{\nu}}  &0 &0 \cr
0 & 0 & \sqrt{[2]_{\nu} [2]_{\nu}} &0 \cr
0 & 0 & 0 & \sqrt{[3]_{\nu} [1]_{\nu}}  \cr
0 & 0 &0 & 0 \cr
\end{pmatrix}\!,
&& 
J_0  = 
\begin{pmatrix} 3/2 & 0 &0 &0 \cr
0 & 1/2 &0 &0 \cr
0 & 0 & -1/2 &0  \cr
0 & 0 &0 & -3/2 \cr
\end{pmatrix}\!, \ \ \ \ 
\nonumber  \\
\hskip-0.3cm 
J_-  = 
\begin{pmatrix} 0 & 0  &0 &0 \cr
\sqrt{[3]_{\nu} [1]_{\nu}} & 0 & 0 & 0 \cr
0 & \sqrt{[2]_{\nu} [2]_{\nu}} & 0 & 0 \cr
0 & 0 & \sqrt{[3]_{\nu} [1]_{\nu}} & 0 \cr
\end{pmatrix}\!,
&&
R_J   = \begin{pmatrix} 1 & 0 &0 &0 \cr
0 & -1 &0 &0 \cr
0 & 0 & 1 &0  \cr
0 & 0 &0 & -1 \cr
\end{pmatrix}\!.
\end{eqnarray}

\item
{ $j=2$}:
In this case, the matrices appear in the form
\begin{align}
J_+  & = \begin{pmatrix} 0 &  \sqrt{[4]_{\nu} [1]_{\nu}}  &0 &0 &0 \cr
0 & 0 &  \sqrt{[3]_{\nu} [2]_{\nu}}  &0 &0 \cr
0 & 0 & 0 &  \sqrt{[3]_{\nu} [2]_{\nu}}   &0 \cr
0 & 0 &0 & 0 &  \sqrt{[4]_{\nu} [1]_{\nu}}  \cr
0 & 0 &0 & 0  & 0 \cr
\end{pmatrix},
\nonumber  \\
J_-  & = \begin{pmatrix} 0 & 0 &0 &0 &0 \cr
\sqrt{[4]_{\nu} [1]_{\nu}} & 0 &0 &0 &0 \cr
0 & \sqrt{[3]_{\nu} [2]_{\nu}} & 0 &0  &0 \cr
0 & 0 &\sqrt{[3]_{\nu} [2]_{\nu}} & 0 &0 \cr
0 & 0 &0 & \sqrt{[4]_{\nu} [1]_{\nu}}  & 0 \cr
\end{pmatrix},
\\
J_0  & = \begin{pmatrix} 2 & 0 &0 &0 &0 \cr
0 & 1 &0 &0 &0 \cr
0 & 0 & 0 &0  &0 \cr
0 & 0 &0 & -1 &0 \cr
0 & 0 &0 & 0  &-2 \cr
\end{pmatrix},
\qquad\qquad
R_J = \begin{pmatrix} 1 & 0 &0 &0 &0 \cr
0 & -1 &0 &0 &0 \cr
0 & 0 & 1 &0  &0 \cr
0 & 0 &0 & -1 &0 \cr
0 & 0 &0 & 0  &1 \cr
\end{pmatrix}.
\nonumber 
\end{align}
\end{itemize}

\subsection{The Holstein-Primakoff realization of $sl_{\nu}(2,R)$ algebra}\label{5}

For the ordinary $sl(2,R)$ algebra,  the Holstein-Primakoff  single-mode realization has the form \cite {Holstein}
\be
J_+ = \sqrt{ 2 j -N} a , \qquad J_- = \ad \sqrt{ 2 j -N}, \qquad J_0= j -N.
\ee
The latter is related to the $(2j+1)$-dimensional Fock space spanned by $|n\rangle$, which obeys $N|n\rangle=n|n\rangle, n=0, 1, \dots, 2j$. A comparison with equation \eqref{43} gives
\be
\frac{N_1-N_2}{2} \Longleftrightarrow  j -N, ~~~ R_1=R_2 =R.
\ee
Thus, the algebra  $sl(2,R)$  reads
\be
[J_+, J_-] = 2 J_0 ( 1 + 2 \nu R), \qquad 
[J_0,J_\pm]=\pm J_\pm.
\ee
Now, we can find the Holstein-Primakoff realization of algebra  $sl(2,R)$, which is given by
\begin{align}
J_0  & = j -N, \qquad\qquad\qquad\qquad R= (-1)^N,
\nonumber  \\
J_+  & = \sqrt{ \frac{ ( 2 j -N)(N+1) + \nu ( 1 + 2 j + R ( -1 +2j -2 N))}{ N+1 + \nu ( 1 + R)}}\ a,
\\
J_-  & = \ad\ \sqrt{ \frac{ ( 2 j -N)(N+1) + \nu ( 1 + 2 j + R ( -1 +2j -2 N))}{ N+1 + \nu ( 1 + R)}}.
\nonumber 
\end{align}

\section{The $so_{\nu}(3)$ algebra}\label{6}

Finally, from a basis change for part of the generators which reflects the isomorphism of $so_{\nu}(3)$ and $su_{\nu}(2)$  we comment on the corresponding modified algebra
\be
L_z = J_0, \qquad L_x =\frac{1}{2} ( J_+ + J_-), \qquad L_y = \frac{i}{2} ( J_- - J_+).
\ee
As a result, the algebra $so_{\nu} (3)$  is given by
\begin{align}
[L_z, L_x] & = i L_y , \qquad [L_z, L_y]= -i L_x,
\nonumber  \\
[L_x, L_y] & =2 L_z  + 2 \nu P + 2 \nu ( 2 \nu +1 ) K,
\nonumber  \\
[K, Q]  & = [K, P] = [K, L_z]=0, \qquad\{ K , L_x\}=\{ K , L_y\}=0,
\\
[Q, P]  & = [ Q, L_z] =0, \qquad \{ Q, L_x \} =\{ Q, L_y \} =0,
\nonumber  \\
[P, L_z ] & =0, \qquad \{ P, L_x  \} = 2i Q L_y, \qquad \{ P, L_y  \} = -2i Q L_x.
\nonumber 
\end{align}

\subsection{ Representation for odd $2j$}
In this case, denoting $R_L = (-1)^{ j - L_z}$, the $so_{\nu}(3)$ algebra has three generators that satisfy:
\be 
[L_z, L_x]  = i L_y , \qquad [L_z, L_y]= -i L_x,
\qquad
[L_x , L_y]   =2 L_z  + 2 \nu  ( 2 \nu + j +1 ) R_L.
\ee

\subsection{ Representation for even $2j$}
Now the algebra $so_{\nu}(3)$ also has three generators:
\be 
[L_z, L_x]   = i L_y , \qquad [L_z, L_y]= -i L_x
\qquad
[L_x , L_y]    =2 L_z ( 1  + 2 \nu   R_L).
\ee

\section{Conclusions}\label{7}

Starting from very recent applications of some concepts from various fields of theoretical physics and quantum technologies, we have developed a new parity-deformed type of Wigner algebras, these being the essential reasons for introducing these new mathematical structures:
\begin{itemize}
\item 
The broad application of parity-deformed systems in quantum information and communication, including the study of coherent and compressed states, cat states, data transfer, and error correction.

\item 
 The application of the Holstein-Primakoff and Jordan-Schwinger realizations in various fields, including quantum communication.

\item 
 The usefulness of $SO(3)$ group representations in the study of multi-qubit systems.

\item 
The experimental limitations and the importance of theoretical calculations in the modeling and preparation of experiments in the areas already mentioned.

\item 
 The application of the groups considered in this work in the study of qutrits.

\item 
The possible application and generalization of traditional light-matter Hamiltonians, including the conventional Rabi model and its recent generalizations. 

\item 
The possible investigation of the two-mode Landau problem in parity-deformed formalism.

\item 
Notable applications of harmonic vibronic models in solid-state, molecular and chemical physics and their applications in quantum technologies.
\end{itemize}

As a result of this work, and based on the single mode and two mode Wigner algebra, the deformed algebra  $sl(2,R)$  ($sl_{\nu}(2,R)$) and the deformed algebra $so( 3)$  ($so_{\nu}(3)$) were constructed.
First, the single-mode Wigner algebra and two types of its realizations were analyzed: one based on the monomial basis and another based on the graded quasi-polynomial basis $Z_2$.
The single-mode Wigner algebra was then extended to the two-mode case, which was used to obtain the algebra $sl_{\nu}(2,R)$. 
The following results were found: (a) for $j=1/2$  we have the same algebra as $sl(2,R)$, (b) the case $j=1$ produces the quadratic algebra and (c) the cases $ j=3/2, 2, 5/ 2, \dots$ result in the algebra $sl_{\nu}(2,R)$, which is neither the algebra $su(2)$ nor the quadratic algebra.
The Holstein-Primakoff realization of this algebra was also discussed, and finally the algebra $so_{\nu}(3)$  was derived from the isomorphy with algebra $su_{\nu}(2)$.

Having prepared the required deformed basis, our next goal is to generalize the present work and investigate its consequences on the temporal evolution of certain related quantum systems, including the analysis of the fidelity of entanglement transfer as well as the  entanglement transfer in the so-called  ${J_x}$  photonic networks.
We hope that the present formulation helps generalizing the conventional Rabi model in a novel manner, a topic we are currently working on.

\section*{Aknowledgments}

This research was supported by the Q-CAYLE project, funded by the European Union-Next Generation UE/MICIU/Plan de Recuperacion, Transformacion y Resiliencia/Junta de Castilla y Leon (PRTRC17.11), and also by RED2022-134301-T and PID2023-148409NB-I00, financed by MI-CIU/AEI/10.13039/501100011033.  
Financial support of the Department of Education of the Junta de Castilla y León and FEDER Funds is also gratefully acknowledged (Reference: CLU-2023-1-05). The authors are grateful to the referees for their critical comments and suggestions and also to Prof. G. Junker, Dr. I.A. Bocanegra-Garay and Dr. D.G. Bussandri for their kind assistance on the manuscript.

\end{document}